\documentstyle[prl,aps]{revtex}

\input epsf
\begin{document}
\draft \twocolumn[\hsize\textwidth\columnwidth\hsize\csname
@twocolumnfalse\endcsname

\title{Novel continuum modeling of crystal surface evolution}
\author{Navot Israeli\cite{NavotEmail} and Daniel Kandel\cite{DanielEmail}}
\address{Department of Physics of Complex Systems, Weizmann Institute
of Science, Rehovot 76100, Israel} \maketitle

\begin{abstract}
We propose a novel approach to continuum modeling of the dynamics of
crystal surfaces. Our model follows the evolution of an ensemble of
step configurations, which are consistent with the macroscopic surface
profile. Contrary to the usual approach where the continuum limit is
achieved when typical surface features consist of many steps, our
continuum limit is approached when the number of step configurations of
the ensemble is very large. The model can handle singular surface
structures such as corners and facets. It has a clear computational
advantage over discrete models.
\end{abstract}

 ]

The behavior of classical physical systems is typically described in
terms of equations of motion for discrete microscopic objects (e.g.\
atoms). In many cases, the behavior of such systems is smooth when
observed on macroscopic length and time scales. It is useful to
describe these systems in terms of continuum, coarse-grained models,
which treat the dynamics of the macroscopic, smoothly varying, degrees
of freedom rather than the microscopic ones. Such models are more
amenable to analytical treatments and have enormous computational
advantages over their discrete counterparts.

Many physical systems exhibit a macroscopically smooth behavior
everywhere, except in small regions of space where their behavior is
singular. Examples are dislocations in a crystal, cracks in crystalline
material, facet edges on crystal surfaces, etc. These singular regions
are of interest because they frequently drive the dynamics of the whole
system. It is an interesting and important challenge to develop
continuum descriptions of such systems, since standard phenomenological
continuum models completely fail in the singular regions.

In this Letter we address the above problem in the context of the
dynamics of crystal surfaces. Below the roughening temperature the
evolution of these surfaces proceeds by the nucleation, flow and
annihilation of atomic steps. These steps originate either from a
miscut of the initial surface with respect to a high symmetry plane of
the crystal or as the boundaries of islands and vacancies which
nucleate on the surface during morphological evolution (we ignore screw
dislocations as sources of steps). The steps are separated by terraces,
which are parallel to a high symmetry crystal orientation. Each step
consists of atomic kinks separated by straight portions along
closed-packed orientations. However, on length scales larger than the
typical distance between kinks the step is a smooth line. Thus the
evolution of the surface corresponds to the motion of discrete smooth
lines.

One can model step motion by solving the diffusion problem of adatoms
on the terraces with boundary conditions at step edges. These
conditions account for attachment and detachment of atoms to and from
step edges. The local flux of adatoms at step edges and the resulting
step velocities can then be calculated. This approach was introduced
long ago by Burton, Cabrera and Frank \cite{BCF}, and was further
developed by other authors \cite{reviews}. Given the arrangement of few
neighboring steps, these models allow a calculation of the normal
velocity of a step. The number $M$ of neighboring steps  on each side
required for this calculation depends on the step kinetics and on the
interaction range between steps (in the examples below $M=2$). Step
models are capable of describing surface evolution on the mesoscopic
scale with significant success \cite{Williams,Blakely}. Such models
pose a serious challenge for numerical computations, and can be solved
only for small systems.

Several attempts were made to construct continuum models of stepped
surfaces
\cite{Mullins,OzdemirZangwill,Nozieres,LanconVillain,Uwaha,%
ChameRoussetBonzelVillain,HagerSpohn,sine_scaling,cone,1D_scaling,%
BonzelPreussSteffen,BonzelPreuss,Murty}, to study their large scale
properties. The general idea behind these attempts is that step flow
can be treated continuously in regions where every morphological
surface feature is composed of many steps. If we label surface steps by
the index $n$, the continuum limit in these models is obtained by
taking $n$ to be continuous. The outcome of these attempts are partial
differential equations for surface evolution. Such continuum models are
fairly successful in describing the evolution of smooth surfaces of
very simple morphology. However, below the roughening temperature,
crystal surfaces have singularities in the form of corners and
macroscopic facets. The latter are a manifestation of the cusp
singularity of the surface free energy at high symmetry crystal
orientations. The assumption that every surface feature is composed of
many steps clearly breaks down on macroscopic facets where there are no
steps at all. Thus, existing continuum models fail near singular
regions.

A possible resolution of this problem is to solve the continuum model
in non-singular regions and use matching conditions at the singular
facet edges \cite{HagerSpohn,sine_scaling,cone,1D_scaling}. This is a
good solution in extremely simple cases (see, e.g. \cite{1D_scaling}),
but generally it is not possible to derive the appropriate matching
conditions from the microscopic models \cite{sine_scaling,cone}. As a
result this approach is not very useful.

Another suggested solution is to round the surface free energy cusp
\cite{BonzelPreussSteffen,BonzelPreuss,Murty}. The resulting continuum
models replace facets by relatively flat but analytic regions. As
explained below, this approach does not correctly reproduce the
evolution of facet edges. Thus, all existing continuum models fail in
the description of the evolution of surfaces with singularities. This
failure is both conceptual and practical, in the sense that it leads to
qualitatively as well as quantitatively incorrect results.

\begin{figure}[h]
\centerline{ \epsfxsize=78mm \epsffile{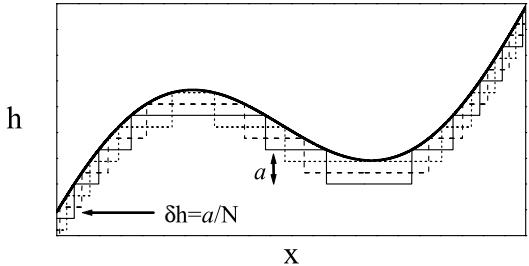}} \vspace{-5mm}
\caption{A schematic illustration of the ensemble of configurations
whose upper envelope defines the configurational continuum limit.}
\label{ensemble}
\end{figure}

We now propose a conceptually new definition of the continuum limit,
which we term {\em configurational continuum}. This definition allows
construction of continuum models, free of all the limitations of
standard continuum models. Our key observation is that a continuous
surface height profile can be represented by many similar, but not
identical, step configurations. Figure \ref{ensemble} is a
demonstration of this point for a one dimensional surface. It shows a
continuous height function, $h(x)$, as a function of position $x$
(thick solid line), and three valid discrete representations of this
profile as step configurations. In configurational continuum we define
the continuous height profile as the upper envelope of the discrete
height functions of an {\em ensemble} of many such step configurations.
To construct these configurations, let $a$ be the height of a single
step and $N$ the number of configurations in the ensemble. We construct
the ensemble so that the height difference $\delta h$, between two
adjacent configurations is $a/N$, as depicted in Fig.\ \ref{ensemble}.
The configurational continuum limit is obtained when $N \rightarrow
\infty$. The generalization to higher dimensions is straightforward.

The dynamics of the continuum model is as follows. Each step
configuration of the ensemble evolves according to the microscopic
dynamics. As a result, the envelope of discrete height functions
changes with time, thus defining the evolution of the continuous height
function.

The crucial assumption hidden in this definition of the continuum model
is that our construction leads to a mathematically well-defined height
function at all times. When does this assumption hold? Consider two
initially similar configurations of the ensemble. Our continuum limit
is well-defined provided these two configurations have similar
microscopic dynamics and hence remain similar at later times
\cite{overhangs}. This assumption has to hold in the standard continuum
definition as well. If two initially similar configurations evolve very
differently, one must follow the specific microscopic configuration of
interest.

We now derive the evolution equation for the continuous height of a two
dimensional surface, $h(\vec{r},t)$, at position $\vec{r}$ and time
$t$. Here $\vec{r}$ is a vector in the $xy$ plane, defined as the high
symmetry crystal plane parallel to the terraces. $h$ changes with time
due to the flow of steps through $\vec{r}$, and due to nucleation and
annihilation of steps. At this stage we disregard nucleation processes
and include them later. First, we consider positions, which are not
local extrema of the height profile. It is obvious from the
construction of the configurational continuum, that for each point
$\vec{r}$ there is exactly one configuration in the ensemble, which has
a step that passes through $\vec{r}$ at the time $t$. That step lies
along the unique equal-height contour line, which passes through
$\vec{r}$. As is demonstrated in Fig.\ \ref{ensemble}, the exact
positions of neighboring steps in the {\em same} configuration can be
calculated from the knowledge of $h(\vec{r},t)$, and the fact that in
this configuration there is a step at $\vec{r}$. Hence, we can use the
discrete step model to calculate the normal velocity of the step at
position $\vec{r}$ at time $t$. Let us denote this velocity by
$\vec{v}_s(\vec{r},t)$. Note that at different positions, $\vec{v}_s$
is the normal velocity of steps which belong to {\em different}
configurations in the ensemble.

Next we define a two dimensional unit-vector $\hat{n}(\vec{r},t)$ in
the $xy$ plane, which is normal to the contour line at $\vec{r}$, and
points in the direction opposite to $\vec{v}_s(\vec{r},t)$. This allows
us to define the directional gradient
\begin{equation}
\nabla h_{\hat{n}}(\vec{r},t)\equiv
\hat{n}(\vec{r},t)\lim_{\epsilon\rightarrow
0_+}\frac{h(\vec{r}+\epsilon\hat{n},t)-h(\vec{r},t)}{\epsilon}
\label{gradient}\;,
\end{equation}
where the directionality is needed in order to deal with corners and
facet edges. The gradient is in the direction from which steps flow
towards the point $\vec{r}$. This is useful for the calculation of the
current of steps arriving at $\vec{r}$:
\begin{equation}
J(\vec{r},t)=\frac{N}{a}\left|\nabla h_{\hat{n}}(\vec{r},t)
\cdot\vec{v}_s(\vec{r},t)\right|\;, \label{current}
\end{equation}
where we have used the fact that the local step density is
$\left|\nabla h_{\hat{n}}(\vec{r},t)\right|N/a$. Note that $J$ is the
current of steps belonging to {\em all} configurations in the ensemble,
and not to one particular configuration. Since each step (from any
configuration), which passes through $\vec{r}$ changes the height of
the ensemble envelope by $a/N$ (see Fig.\ \ref{ensemble}), the
continuous height profile obeys the evolution equation
\begin{equation}
\frac{\partial h(\vec{r},t)}{\partial t}=-\nabla
h_{\hat{n}}(\vec{r},t)\cdot\vec{v}_s(\vec{r},t)\;.
\label{dynamic_equation}
\end{equation}

The above derivation of the evolution in the continuum is not valid at
local extrema of the surface, because generally one cannot define a
unique equal-height contour line which passes through such a point. To
avoid the problem, we define $\partial h /
\partial t$ at local extrema as the limit of the height time derivative
as one approaches these points. This is justified, since there are no
microscopic realizations of the surface with steps exactly at the local
extrema.

At this point we emphasize that the configurational continuum evolution
is formally identical to the evolution of the discrete step model. This
statement is almost trivial, since by definition Eq.\
(\ref{dynamic_equation}) follows the envelope defined by the ensemble
of configurations, and each configuration evolves with step velocities
calculated from the discrete step model. Thus Eq.\
(\ref{dynamic_equation}) is exact. Similarly to other continuum models
it is solved numerically by discretization of space, which is the only
approximation involved in such solutions.

\begin{figure}[h]
\centerline{ \epsfxsize=78mm \epsffile{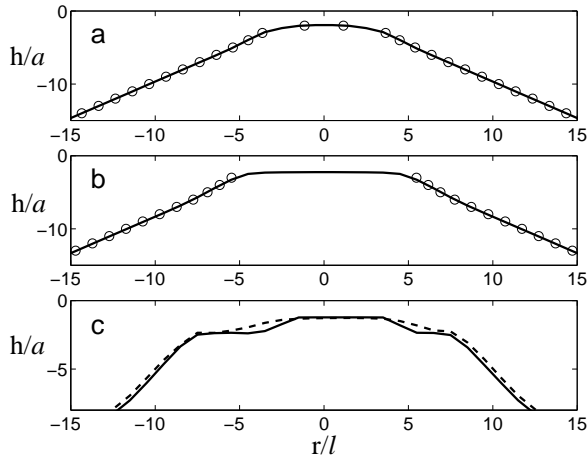}} \vspace{1mm}
\caption{Evolution of a crystalline cone with growth flux. The height
(in units of $a$) is shown vs.\ the radius (in units of the initial
terrace width, $l$). The lines show cross sections of the two
dimensional solution of Eq.\ (\ref{dynamic_equation}). Circles show the
surface evolution according to the one dimensional step flow model. The
initial shape is an exact cone, and (a) and (b) show the generation and
evolution of a facet at the top at later times. (c) shows an island
which nucleated (solid) and grew (dashed) on the top facet.}
\label{growing_cone}
\end{figure}

What is the relation between configurational continuum and other
existing continuum models? We will show elsewhere that in regions where
$h(\vec{r},t)$ is a smooth function of $\vec{r}$, the evolution
equation (\ref{dynamic_equation}) reduces to the differential equation
of standard continuum models. However, near corners or facets
(\ref{dynamic_equation}) accounts for the discrete nature of steps and
cannot be written as a differential equation, due to the singular
nature of the profile. Contrary to other continuum models,
configurational continuum is exact even at singular points and is
capable of describing both smooth and singular behaviors without
artificial smoothing or matching conditions. To the best of our
knowledge, configurational continuum is the only continuum model
capable of correctly treating the dynamics of singular lines
\cite{levelset}.

The fact that no matching conditions are needed in configurational
continuum is truly important for possible generalizations. A matching
condition approach has been used in other areas, such as plasticity of
crystals. The idea there is to solve continuum elasticity far from
dislocations and a discrete model near dislocations. The inaccurate
matching conditions between the different regions lead in this case to
artificial reflection of phonons from the boundaries between the
regions \cite{yip}. A proper generalization of configurational
continuum to atomistic systems may eliminate these artificial phonon
reflections.

In terms of computational efficiency, our model is comparable to
standard continuum models. The computational advantage over discrete
models arises from the fact that in macroscopically smooth regions the
density of steps is much larger than the needed density of
discretization points. Some care should be taken near singular lines,
and discretization there should be dense enough to allow an accurate
interpolation of the singular profile. This, however, should not affect
the overall efficiency, and when most of the surface is smooth the
number of steps needed to be simulated is usually orders of magnitude
smaller than the number of steps in the system.

We now apply the configurational continuum approach to a few simple
cases in order to demonstrate its validity. To this end, we consider
the evolution of a conic structure which consists of circular
concentric steps. The adatom concentration on the terraces can be
integrated out due to the radial symmetry of the system, yielding a set
of coupled non-linear differential equations for the evolution of step
radii \cite{cone}. We have solved this discrete model numerically with
and without growth flux. In the absence of flux, the cone decays by
step flow. During the decay, a facet develops at the top of the cone,
and its radius grows as $t^{1/4}$. With flux, the cone grows except at
the peak which initially decays and then saturates. A facet forms at
the peak after saturation.

The continuum model we solved for this example is a two dimensional
model, which can, in principle, develop non radially symmetric
morphology. The microscopic dynamics we used was a two dimensional
generalization of the microscopic equations for step radii of the
discrete one dimensional model. Any microscopic dynamics, such as a
full solution of the diffusion equation on each terrace, can be used in
the framework of configurational continuum. For the sake of
demonstrating the validity of our approach the simple dynamics we chose
is sufficient.

Figures \ref{growing_cone}(a) and (b) show a comparison between a
numerical solution of the discrete step model with growth flux and a
cross section of the two dimensional surface obtained from the solution
of our continuum model at various times. The agreement is quite
impressive. In particular, the width and height of the facet obtained
from the continuum model are equal (within numerical accuracy) to those
of the discrete model at all times. A similar agreement between the two
models is achieved in the absence of growth flux. For comparison, we
also investigated the dynamics of the decaying cone using the continuum
model of \cite{Murty}. Within this model, the cusp in the free energy
is rounded, and the evolution is studied with various degrees of
rounding. We found that even in the limit of small rounding the height
profile is very different from the profile of the discrete model.
Although a facet does develop, its size is about 40\% larger than the
facet of the discrete model at all times. In addition, the shapes of
the height profile differ qualitatively.

Island nucleation can also be included in our continuum model. To this
end we add a microscopic nucleation scheme, which defines the
nucleation probability at any given position on the surface, given a
specific microscopic step configuration. Within our continuum approach,
the nucleation probability at a point on the continuous surface is the
ensemble average of the microscopic nucleation probabilities at this
point. For demonstration purposes we included the simple scheme where
the nucleation probability in any given step configuration is
proportional to the square of the local adatom concentration. In fact,
island nucleation is already included in the simulations presented in
Fig.\ \ref{growing_cone}. Nucleation events occur on the top facet once
it becomes large enough, but there is hardly any nucleation on the
finite slope regions, because the terraces there are very narrow.
Figure \ref{growing_cone}(c) shows an island on the top facet, which
nucleated and started growing. Vacancy nucleation during evaporation
can be included in a similar manner.

\begin{figure}[h]
\centerline{ \epsfxsize=78mm \epsffile{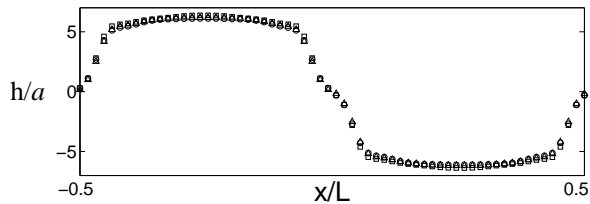}} \vspace{1mm}
\caption{Data collapse for week repulsive interactions between steps.
Shown are cross sections (along the $x$ direction at $y=-L/4$) of
bidirectional sinusoidal profiles with different wave lengths after a
relaxation time of $t_0 L^3$. Wave lengths shown are $L=64$ (circles),
$128$ (squares) and $256$ (triangles).}
 \label{G_egg}
\end{figure}
We now turn to the more demanding example of bidirectional sinusoidal
grating relaxation. Here the initial surface height profile of wave
length $L$ is given by $h_L(\vec{r},t=0)=h_0 \sin\left(2\pi x/L\right)
\sin\left(2 \pi y/L\right)$. The relaxation of this profile towards a
flat surface was studied by Rettori and Villain \cite{RettoriVollain},
who gave an approximate solution to a step flow model, in the limit
where the interaction between steps can be neglected in comparison with
the step line tension. The decay of bidirectional sinusoidal profiles
was also studied numerically \cite{Murty}. We now apply our model to
this problem assuming diffusion limited kinetics without deposition
flux.

For weak interactions between steps, the surface height evolves
according to $\partial h_L/\partial t\propto \nabla^2 \kappa \sim
L^{-3}$, where $\kappa$ is the step curvature. We therefore expect the
following scaling law: $h_L(\vec{r},t)=h_{L=1}(\vec{r}/L,t/L^3)$.
Figure \ref{G_egg} shows the data collapse of cross sections of
profiles from the configurational continuum model. The different
symbols correspond to different wave lengths at time $t=t_0 L^3$ for a
fixed $t_0$. The quality of the data collapse shows that the scaling
scenario holds very accurately. Note that large facets have developed
at the surface extrema, and they are connected by very steep slopes.
This shape does not agree with Rettori and Villain's heuristic argument
\cite{RettoriVollain}, which predicts that facets appear also near
$h=0$ lines. Nevertheless, their prediction that the decay time of the
height profile goes as $L^3$ is in agreement with both the scaling law
and our numerical solutions. Our simulations show that for much
stronger repulsive interactions between steps, the scaling law does not
hold.


\begin{thebibliography}{99}
\bibitem{overhangs}
Note that the height is not well-defined if overhangs are generated. In
this case it is possible to generalize the continuous function to
appropriately describe the overhangs.
\end{thebibliography}
\end{document}